\makeatletter
\newcommand{\dontusepackage}[2][]{%
  \@namedef{ver@#2.sty}{9999/12/31}%
  \@namedef{opt@#2.sty}{#1}}
\makeatother
\dontusepackage{subfigure}

\documentclass[]{article}

\usepackage{lmodern}
\usepackage{amssymb,amsmath}
\usepackage{ifxetex,ifluatex}
\usepackage[usenames,dvipsnames]{color}
\usepackage{fixltx2e} 
\ifnum 0\ifxetex 1\fi\ifluatex 1\fi=0 
  \usepackage[T1]{fontenc}
  \usepackage[utf8]{inputenc}
\else 
  \ifxetex
    \usepackage{mathspec}
    \usepackage{xltxtra,xunicode}
  \else
    \usepackage{fontspec}
  \fi
  \defaultfontfeatures{Mapping=tex-text,Scale=MatchLowercase}
  
\fi
\IfFileExists{upquote.sty}{\usepackage{upquote}}{}
\IfFileExists{microtype.sty}{%
\usepackage{microtype}
\UseMicrotypeSet[protrusion]{basicmath} 
}{}
\usepackage[margin=1.0in,bottom=1.5in]{geometry}
\usepackage[]{natbib}
\usepackage{listings}
\usepackage{graphicx}
\makeatletter
\def\maxwidth{\ifdim\Gin@nat@width>\linewidth\linewidth\else\Gin@nat@width\fi}
\def\maxheight{\ifdim\Gin@nat@height>\textheight\textheight\else\Gin@nat@height\fi}
\makeatother
\setkeys{Gin}{width=\maxwidth,height=\maxheight,keepaspectratio}
\usepackage{caption}
\usepackage{float}
\ifxetex
  \usepackage[setpagesize=false, 
              unicode=false, 
              xetex]{hyperref}
\else
  \usepackage[unicode=true]{hyperref}
\fi
\hypersetup{breaklinks=true,
            bookmarks=true,
            pdfauthor={},
            pdftitle={3D seismic survey design by maximizing the spectral gap},
            colorlinks=true,
            citecolor=black,
            urlcolor=blue,
            linkcolor=black,
            pdfborder={0 0 0}}
\title{3D seismic survey design by maximizing the spectral gap}
\author{Yijun Zhang\textsuperscript{1,*}, Ziyi Yin\textsuperscript{1,*}, Oscar
López\textsuperscript{2}\\Ali Siahkoohi\textsuperscript{3}, Mathias
Louboutin\textsuperscript{1}, and Felix J.
Herrmann\textsuperscript{1}\\[2\baselineskip]\textsuperscript{1} Georgia
Institute of Technology\\\textsuperscript{2} Florida Atlantic
University\\\textsuperscript{3} Rice University\\\textsuperscript{*}
contributed equally\\Correspondence to ziyi.yin@gatech.edu\\}
\date{}

\begin{document}
\maketitle
\begin{abstract}
The massive cost of 3D acquisition calls for methods to reduce the
number of receivers by designing optimal receiver sampling masks. Recent
studies on 2D seismic showed that maximizing the spectral gap of the
subsampling mask leads to better wavefield reconstruction results. We
enrich the current study by proposing a simulation-free method to
generate optimal 3D acquisition by maximizing the spectral gap of the
subsampling mask via a simulated annealing algorithm. Numerical
experiments confirm improvement of the proposed method over receiver
sampling locations obtained by jittered sampling.
\end{abstract}

\section{Introduction}\label{introduction}

Acquiring fully sampled seismic data is expensive in practice. Thanks to
recent advancements in compressive sensing, seismic data can be randomly
sampled along spatial coordinates to improve acquisition efficiency
\citep{candes2006robust, mosher2014increasing, kumar2015efficient, chiu20193d}.
The subsampled data can be consequently processed by wavefield
reconstruction techniques to recover the fully sampled data
\citep{hennenfent2008simply, kumar2015efficient, zhang2020wavefield}.
While uniform random sampling \citep{candes2010power, candes2012exact}
and jittered subsampling schemes, which control the maximum gap size in
subsampled data \citep{herrmann2008non}, are easy to generate, they may
not be optimal and can be improved through the resolution of specific
optimization problems \citep{mosher2014increasing, manohar2018data}.
Notably, a mutual coherence-based global optimization scheme, initially
proposed by \citet{mosher2014increasing}, \citet{li2017compressive}, and
later expanded upon by \citet{titova2019mutual}, has been developed to
enhance reconstruction quality. \citet{mosher2014increasing} proposed a
simulation-based acquisition design method based on the advancements in
compressive sensing to search for optimized sampling schemes. Similarly,
\citet{guo2020data} optimized time-lapse seismic acquisition using prior
seismic data \citep{manohar2018data}. \citet{guo2023optimal} also
proposed a simulation-based survey design method based on reinforcement
learning. These methods have shown promising results, but determining
the optimal source-receiver layout through combined wavefield
simulations and recoveries is computationally expensive and requires
detailed information of the seismic data, making it infeasible for
optimizing for source and receiver locations in a large-scale 3D survey.

Matrix completion is a computationally efficient technique used to
reconstruct fully sampled data from sparsely sampled seismic data
\citep{kumar2015efficient}. In matrix completion theory, the spectral
gap, which measures the connectivity of the graph in expander graph
theory, is employed to predict and partially quantify the quality of the
matrix completion result solely based on the binary subsampling mask
\citep{bhojanapalli2014universal}. \citet{lopez2023spectral} further
confirms that the success of seismic wavefield reconstruction through
universal matrix completion can be predicted based on the ratio of the
first two singular values of the binary subsampling masks.

While recent work by \citet{lopez2023spectral} eliminates the need for
multiple expensive wavefield reconstructions with different sampling
mask candidates, it does not yet provide a method to generate sampling
masks that maximize the spectral gap. \citet{zhang2022simulation}
introduced a practical algorithm utilizing simulated annealing
\citep{henderson2003theory} to generate acquisition geometries with
small spectral gap ratios, and this algorithm was further extended to
time-lapse seismic acquisition by \citet{zhang2023otl}. However, the
application of this technique to 3D seismic cases has not been explored
yet. In this expanded abstract, we will introduce a practical algorithm
aimed specifically at minimizing the spectral gap ratio of the binary
subsampling mask corresponding to a 3D seismic survey.

We organize this expanded abstract as follows. First, we present the
proposed optimization problem aimed at maximizing the spectral gap ratio
of 3D subsampling masks. Next, we introduce an intriguing property of
spectral gap ratio of the binary masks for the unique case of 3D seismic
surveys. Thanks to this property, we are able to reduce the
computational cost of our algorithm. Finally, we demonstrate numerical
experiments conducted on the synthetic 3D Compass dataset
\citep{jones2012building} and showcase an enhancement in recovery
quality by juxtaposing two wavefield reconstruction results where the
receiver locations are obtained by the jittered subsampling method
\citep{hennenfent2008simply} and our proposed method.

\section{Methodology}\label{methodology}

The quality of seismic wavefield reconstruction through universal matrix
completion \citep{bhojanapalli2014universal} can be predicted by
evaluating the ratio between the first two singular values of binary
sampling masks, denoted as
${\sigma_2(\mathbf{M})}/{\sigma_1(\mathbf{M})}\in [0,\, 1]$, where the
binary matrix $\mathbf{M}$ contains $1$'s to indicate sampled data and
$0$'s otherwise. Termed as the spectral gap ratio, this measure offers a
cost-effective means to quantitatively assess recovery quality. A
smaller spectral gap ratio implies enhanced connectivity within the
graphs formed by binary sampling masks, resulting in improved wavefield
recovery \citep{lopez2023spectral}. The spectral gap ratio has proven
effective in predicting and enhancing 2D wavefield reconstruction
performance \citep{zhang2022simulation} and has been successfully
applied to time-lapse survey design \citep{zhang2023otl}. In this
expanded abstract, we specifically consider 3D survey design where
receivers are missing and sources are fully sampled. We propose a cheap
algorithm based on SGR minimization to optimize sparse geometries for 3D
seismic acquisition.

\subsection{Optimized 3D seismic
acquisition}\label{optimized-3d-seismic-acquisition}

3D wavefield reconstruction based on low-rank matrix completion relies
on the non-canonical Source-X/Receiver-X (columns) Source-Y/Receiver-Y
(rows) organization of the data into a matrix in order to leverage the
inherent low-rank characteristics of seismic data
\citep{kumar2015efficient}. Following this success, we aim to minimize
the SGR of the subsampling mask in the same domain. In this scheme, the
subsampling mask is represented as
$\mathbf {M} \in \{0,1\}^{(N_{sx}\times N_{rx}) \times (N_{sy}\times N_{ry})}$,
where $N_{sx}$ and $N_{sy}$ represent the number of sources along the
$x$ and $y$ coordinates, respectively, and $N_{rx}$ and $N_{ry}$
represent the number of receivers along these coordinates. Motivated by
the success on (time-lapse) 2D seismic survey design reported by
\citet{zhang2022simulation} and \citet{zhang2023otl}, we solve the
following optimization problem to minimize the spectral gap ratio of the
3D acquisition mask:
\begin{equation}
\underset{\mathbf{M}} {\text{minimize}} \quad {\sigma_2(\mathbf{M})}/{\sigma_1(\mathbf{M})}
\quad \text{subject to} \quad \mathbf{M}\in \mathcal{C}.
\label{SG_opt_1V}
\end{equation}
 The objective function is the SGR of the subsampling mask, where
$\sigma_1$ and $\sigma_2$ represent the first and the second singular
values, respectively. In order to ensure the feasibility of optimized
binary masks with a receiver subsampling ratio denoted as
$\rho\in(0,1)$, we incorporate a constraint denoted as
$\mathcal{C}=\bigcap\limits_{i=1}^4 \mathcal{C}_i$. This constraint
encompasses four components: a cardinality constraint defined as
\begin{equation}
\mathcal{C}_1=\{\mathbf{M}\mid\#(\mathbf{M})=\lfloor N_{rx} \times N_{ry}  \times \rho\rfloor \times N_{sx} \times N_{sy} \},
\label{card}
\end{equation}
 a binary mask constraint as
\begin{equation}
\mathcal{C}_2=\{\mathbf{M}\mid \mathbf{M}\in\{0,1\}^{(N_{sx}\times N_{rx}) \times (N_{sy}\times N_{ry})}\},
\label{binary}
\end{equation}
 and constraints, $\mathcal{C}_3$ and $\mathcal{C}_4$, to enforce lower
bounds for the number of subsampling points for each row and column of
the binary subsampling matrix, respectively, in order to avoid missing
row or column in the matrix. They are defined as
\begin{equation}
\begin{split}
\mathcal{C}_3 = \{\mathbf{M}\mid \#(\mathbf{M}_i)\geq m \} \quad \text{and} \quad \mathcal{C}_4 = \{\mathbf{M}\mid \#(\mathbf{M}^j)\geq n \} \\
\text{where} \quad i = 1, \cdots , N_{sx}\times N_{rx} \quad \text{and} \quad j = 1, \cdots , N_{sy}\times N_{ry}.
\end{split}
\label{lowerbound}
\end{equation}
 Here, $\mathbf{M}_i$ and $\mathbf{M}^j$ represent the $i$-th row and
the $j$-th column of matrix $\mathbf{M}$, respectively. $m$ represents
the lower bound for number of subsampling points for each row, and $n$
for each column. These lower-bound constraints ensure that the binary
subsampling matrix does not contain any empty rows or columns. This is
important because having an empty row or column can be detrimental to
the process of matrix completion.

\subsection{An intriguing property of the spectral gap
ratio}\label{an-intriguing-property-of-the-spectral-gap-ratio}

\begin{figure}
\centering
\includegraphics[width=1.000\hsize]{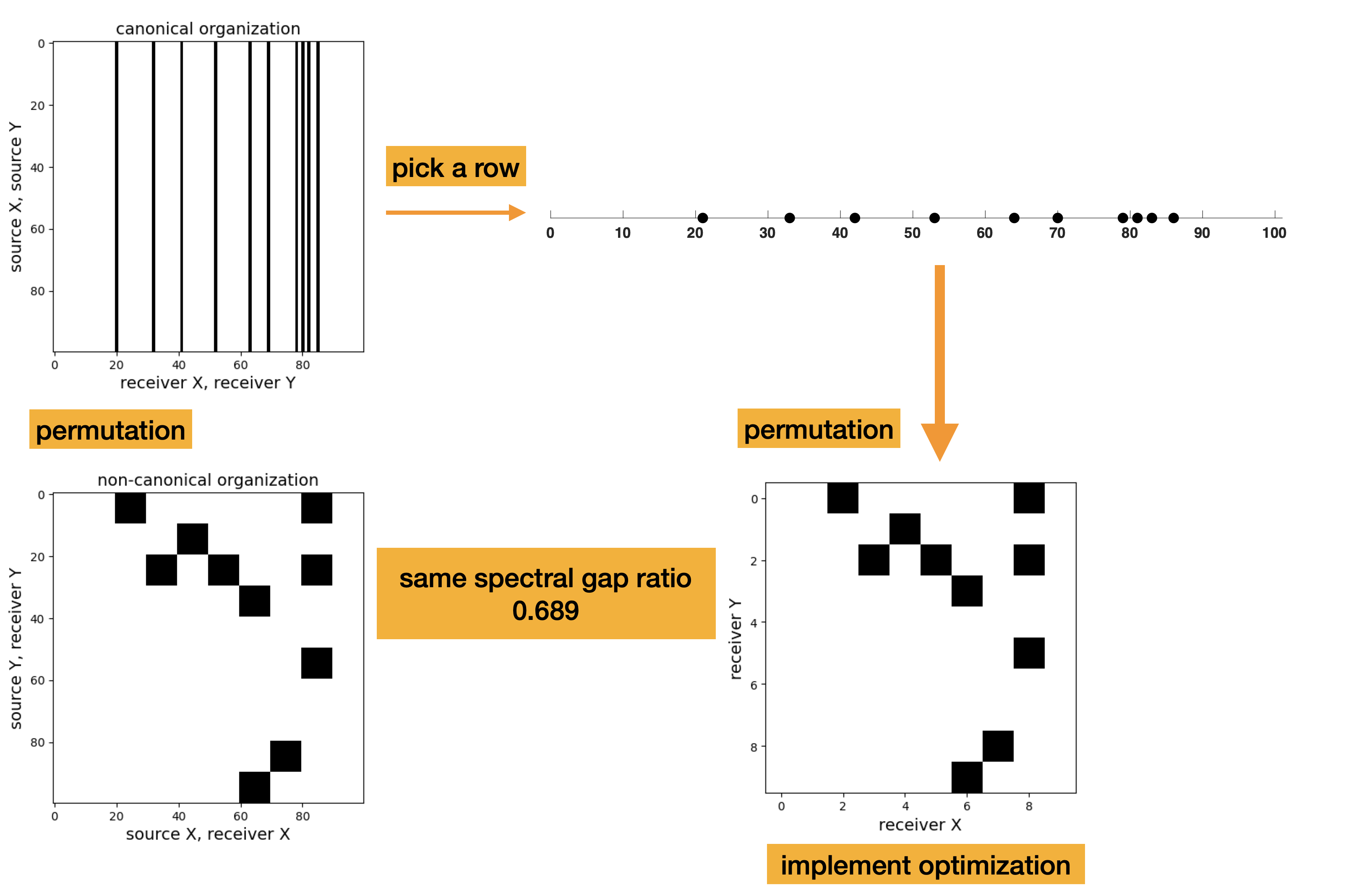}
\caption{Spectral gap ratio of the data matrix in the non-canonical
Source-X/Receiver-X (columns) Source-Y/Receiver-Y (rows) domain is the
same as the spectral gap ratio of the single-source receiver sampling
matrix.}\label{fig1}
\end{figure}

Despite the fact that this proposed optimization could be solved using
simulated annealing \citep{zhang2022simulation, zhang2023otl}, the
binary subsampling matrix $\mathbf{M}$ can still be large for 3D seismic
cases, which potentially slows down the algorithm in practice. To
overcome this problem, we have fortunately discovered that when sources
are fully sampled, each single-receiver block of the global sampling
matrix is either fully sampled or empty depending on whether that
specific receiver is sampled. Consequently, the block structure of the
global matrix leads to the exact same singular values as a single-source
receiver sampling mask, as shown in Figure~\ref{fig1}. We can therefore
optimize a single-source mask to obtain the global optimized mask. The
main computational cost therein is computing the first two singular
values of the receiver sampling mask, which is negligible compared to
approaches that require wave simulations
\citep{mosher2014increasing, guo2023optimal}. The resulting optimal mask
with the lowest SGR indicates the receiver sampling locations that favor
3D wavefield reconstruction via matrix completion in the non-canonical
organization domain.

\subsection{Experiment}\label{experiment}

\begin{figure}
\centering
\includegraphics[width=1.000\hsize]{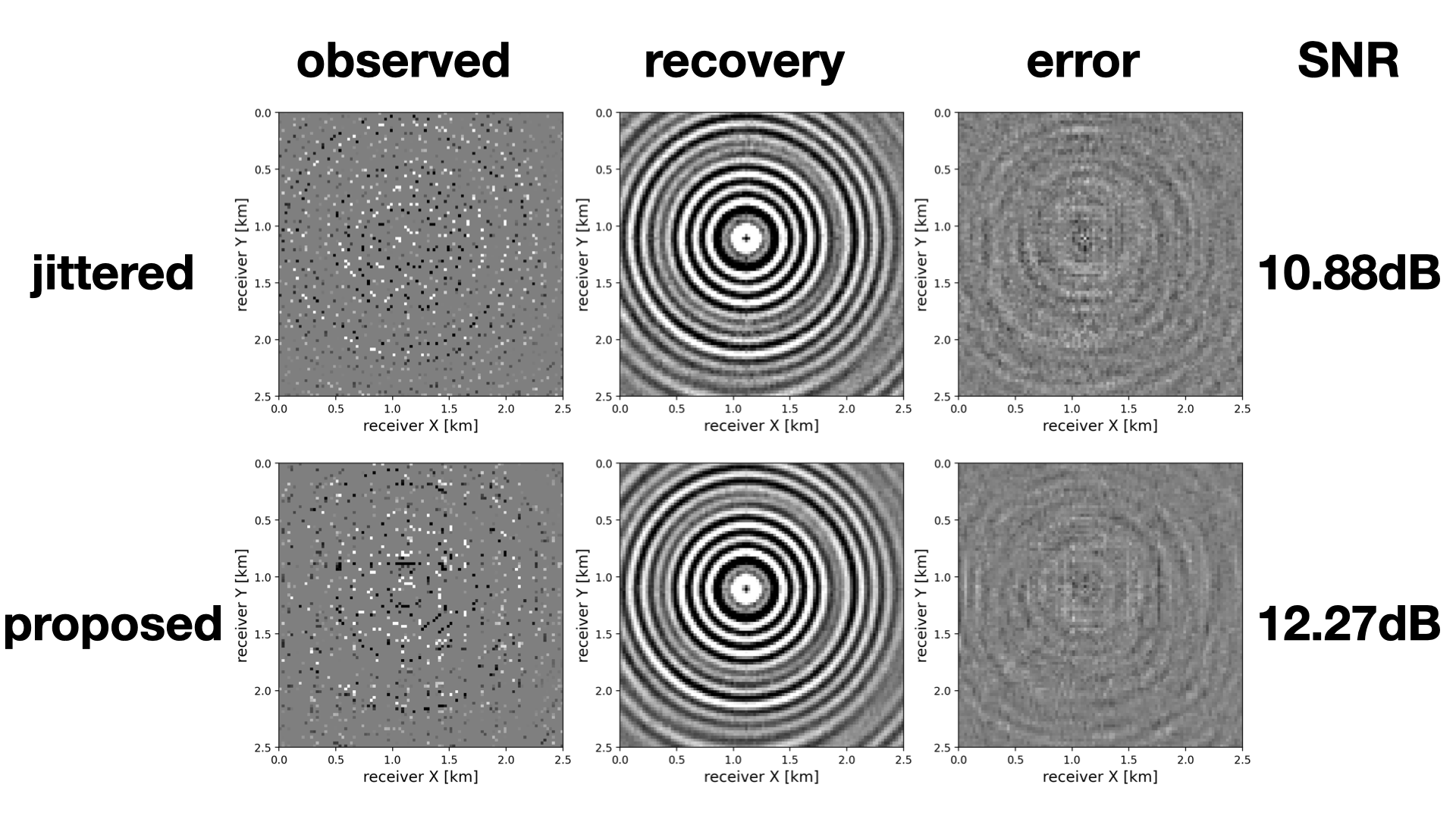}
\caption{Comparison of data reconstruction performance for receiver
locations sampled by the jittered method and the proposed method. There
is about $1.4 dB$ SNR improvement.}\label{fig2}
\end{figure}

We illustrate the efficacy of our method via a numerical experiment on a
simulated 3D marine dataset over the compass model
\citep{jones2012building}. The data volume consists of $501$ time
samples, $1681$ sources and $10,000$ receivers. The distance between the
adjacent sources and receivers are $150 \mathrm{m}$ and $25 \mathrm{m}$
in each direction, respectively, with a time sampling interval of
$0.01 \mathrm{s}$. By using jittered subsampling
\citep{herrmann2010randomized}, we removed $90\%$ of the receivers. This
results in a binary matrix with a SGR of $0.507$ in the non-canonical
domain. After applying simulated annealing algorithm, the SGR of the
mask effectively decreases to $0.328$. To validate the efficacy of our
acquisition design method, we perform data reconstruction on a frequency
slice at $16.8 \mathrm{Hz}$ via weighted matrix completion
\citep{zhang2020wavefield} for the two subsampled datasets with jittered
subsampling mask and the optimized mask. This data reconstruction
process was performed on each dataset individually. The reconstruction
results are shown in Figure~\ref{fig2}, providing evidence for the
efficacy of the acquisition design strategy employed throughout the
study. The reconstruction signal-to-noise ratio (SNR) obtained from data
observed at jittered sampled receiver locations is $10.88 dB$, which is
lower than the reconstruction SNR obtained from data observed at
specified receiver locations obtained by solving
Equation~\ref{SG_opt_1V} ($12.27 dB$). This distinction is crucial, with
an increase of approximately $1.4 dB$ in SNR. This finding supports the
hypothesis that the optimized receiver positions can lead to a superior
seismic survey, ultimately improving the performance of the wavefield
reconstruction. Specifically, this algorithm is computationally
inexpensive compared to simulation-based survey design methods, because
each iteration of this algorithm only needs to calculate the first two
singular values of the receiver sampling mask --- a small
$100 \times 100$ matrix.

\section{Conclusions}\label{conclusions}

Our expanded abstract presents the first numerical case study that
applies spectral gap ratio minimization techniques for 3D seismic
acquisition design. Rather than requiring costly wave simulations, the
proposed method only relies on a single binary matrix optimization which
is computationally inexpensive. Through a representative numerical
experiment conducted on 3D Compass dataset, we conclude that our
proposed method yields an optimal subsampling mask that is highly
suitable for 3D wavefield reconstruction based on matrix completion.
This cheap while effective optimization scheme has the potential to
scale to industry-size 3D survey design problems.

\section{Acknowledgement}\label{acknowledgement}

This research was carried out with the support of Georgia Research
Alliance and partners of the ML4Seismic Center.

\bibliography{abstract}

\end{document}